\documentclass[twocolumn,prb,amsmath,amssymb,showpacs,
superscriptaddress,floatfix]{revtex4}
\usepackage{graphicx,epsfig}

\begin{document}

\title{Electron-electron interactions
in a one-dimensional quantum wire spin filter}

\author{P. Devillard}
\affiliation{Centre de Physique Th\'eorique, Universit\'e de Provence, 
Case 907, 13331 Marseille Cedex 03, France}

\author{A. Cr\'epieux}
\affiliation{Centre de Physique Th\'eorique, Universit\'e de la
M\'editerran\'ee, Case 907, 13288 Marseille Cedex 9, France}

\author{K. I. Imura}
\affiliation{Condensed Matter Theory Laboratory, RIKEN, 2-1 Hirosawa, 
Wako, Saitama 351-0198, Japan} 

\author{T. Martin}
\affiliation{Centre de Physique Th\'eorique, Universit\'e de la
M\'editerran\'ee, Case 907, 13288 Marseille Cedex 9, France}


\begin{abstract}
The combined presence of a Rashba and a Zeeman effect in 
a ballistic one-dimensional conductor generates
a spin pseudogap and the possibility to propagate
a beam with well defined spin orientation.
Without interactions transmission through
a barrier gives a relatively well polarized beam.
Using renormalization group arguments, 
we examine how electron-electron interactions
may affect the transmission coefficient and the
polarization of the outgoing beam.
\end{abstract}


\maketitle


Over the last decade, spintronics\cite{spintronics} has emerged from mesoscopic 
physics and nanoelectronics as a field with implications
in both quantum information theory \cite{quantinfo} and for the storage of 
information. While the charge is routinely manipulated 
in nanoelectronics, the issue here is to exploit 
the spin degree of freedom of electrons. In particular,
spin filters are therefore needed to control the input and 
the output of spintronic devices. A recent proposal \cite{StredaSeba}
explained the operation of a spin filter for a one dimensional wire
under the combined operation of Rashba spin orbit coupling
and Zeeman splitting. Yet, electronic interactions in one 
dimensional wires are known to lead to Luttinger liquid 
behavior and to the renormalization of scattering coefficients. 
It therefore important to inquire about the role
of electronic interactions in  the above mentioned spin filter.
This is aim of the present paper.

A decade ago, 
a transistor  
based on the controlled
precession of the electron spin
due to spin orbit coupling
was proposed \cite{DattaDas}. 
Indeed, the Rashba effect \cite{Rashbaold} in a semiconductor, 
can be modulated by a gate voltage, which 
controls the asymmetry of the potential well
which confines the electrons.
Since this proposal, many spintronic devices
based on the Rashba effect 
have been proposed,  based on a single electron 
picture \cite{Egues,Safarov,Rashbanew, SchliemanLoss}.
At the same time, one-dimensional wires are 
now available experimentally. This has 
motivated several efforts \cite{moroz} to study the 
interplay between the Rashba effect and electron 
interactions in infinite one-dimensional wires.
However, little has been said about Coulomb interactions in 
spintronic devices. Our starting point stems from the fact 
that Rashba devices are gated devices, in which the interaction 
are assumed to be screened and weak. 
We will therefore use a 
perturbative renormalization group treatment of the Coulomb 
interaction in order to address its consequence on the 
spin dependent transmission.  
This approach will be justified by 
estimating the strength of electron-electron interaction 
in single channel semiconductor devices.  
    
\begin{figure}[ht] 
\epsfxsize 6. cm
\epsfysize 4. cm  
\centerline{\epsffile{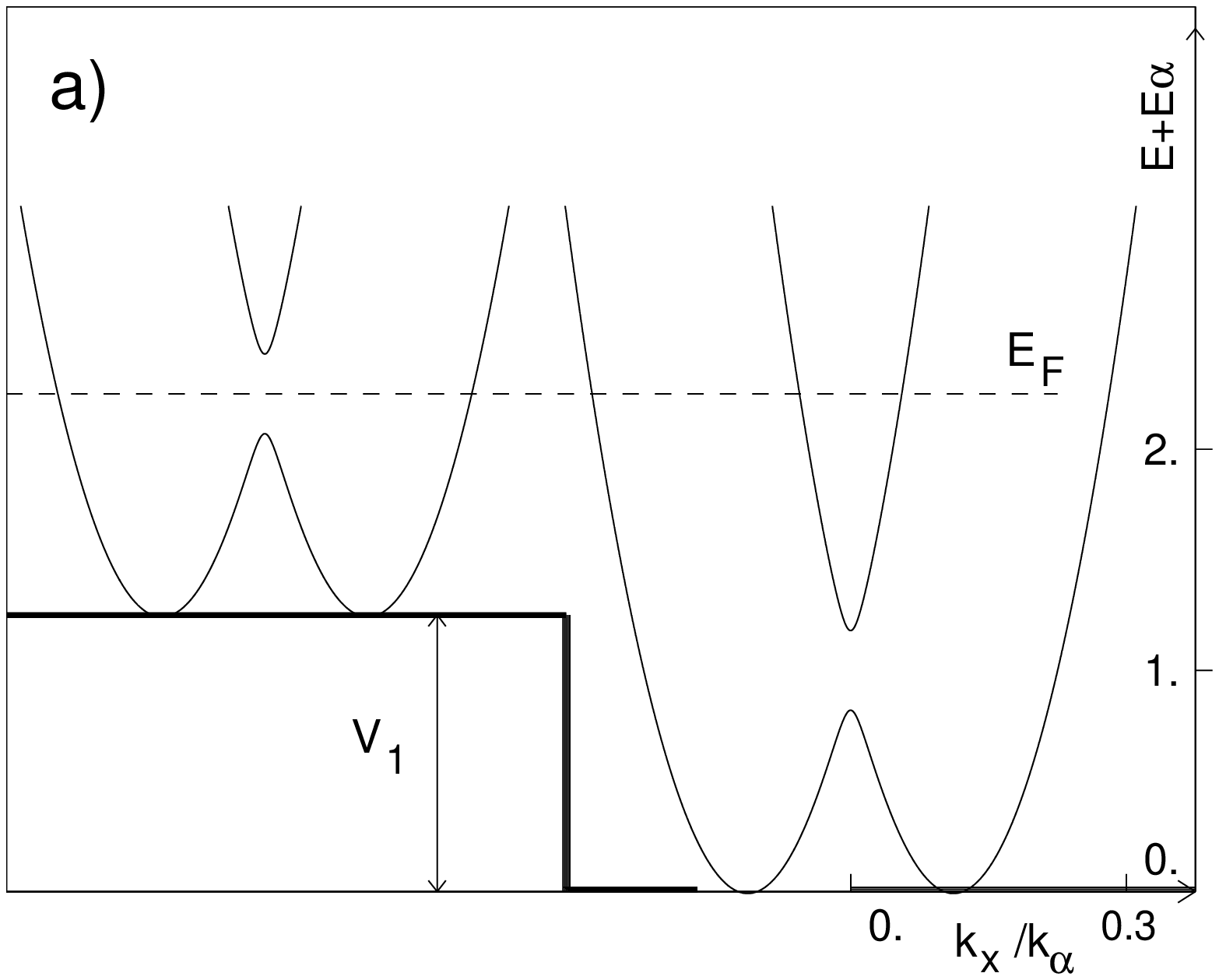}}
\centerline{\epsffile{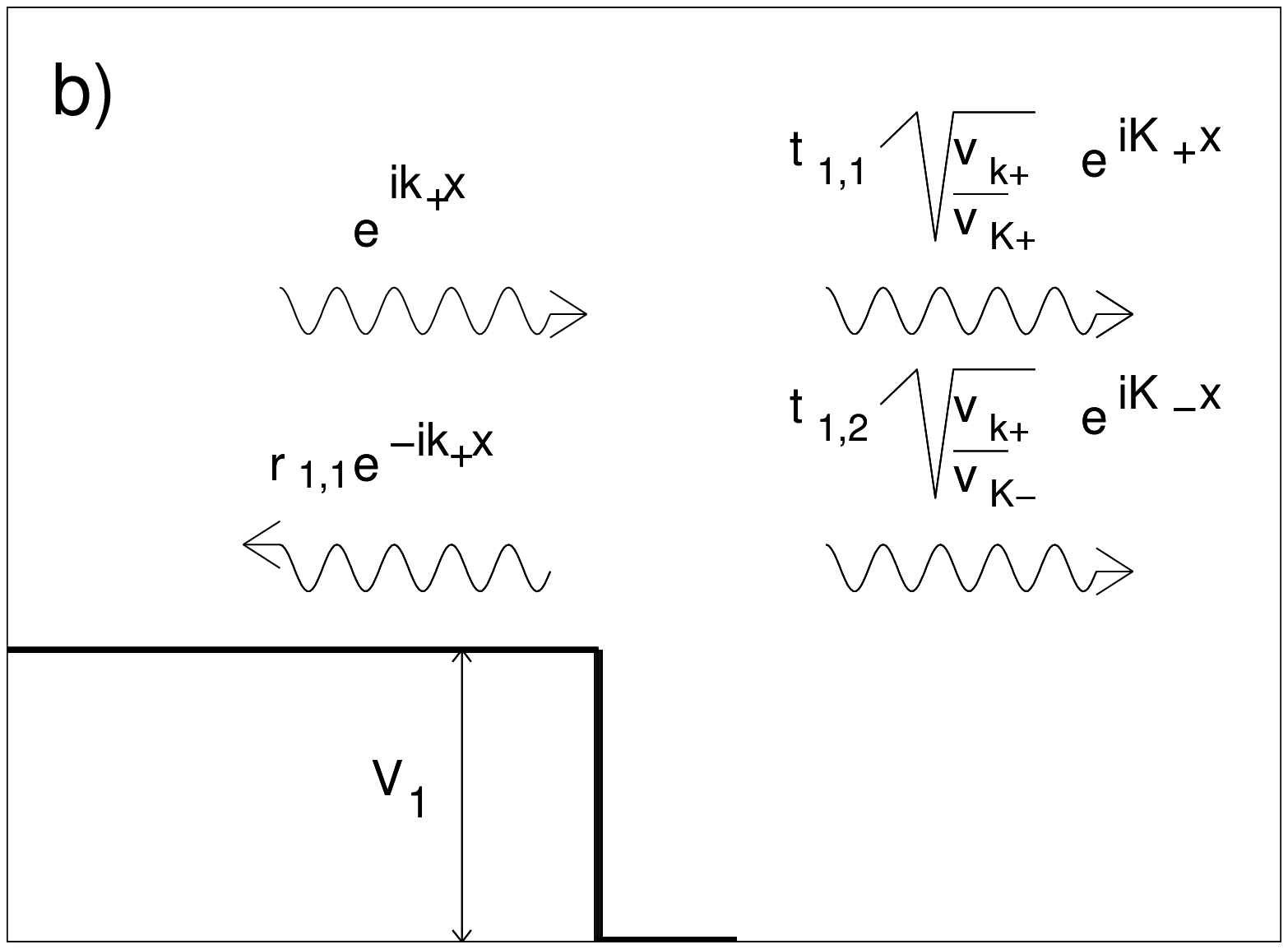}}
\caption{a)  The thin lines denote the energies $E+E_{\alpha}$
of the two Rashba bands as a function of wave-vector ratio
$k_x/k_{\alpha}$ both on the left of the potential step and on the right, 
for $\eta = 0.3$ and $\Gamma=1.5$.
The potential step of height $V_1$ is indicated as a thick line.
The dashed line is the Fermi energy $E_F$. 
b) One of the three scattering states. An incident wave $e^{i k_+ x}$
is injected on the left hand side of the step and 
generates two transmitted waves and one reflected wave.\label{fig1}}
\end{figure}

Here, we consider a narrow ballistic wire which is submitted to spin
orbit coupling with the Rashba term being dominant. In order to
obtain a spin-polarized beam of electrons propagating in one
direction, a small Zeeman field is introduced \cite{StredaSeba}.
By confining in the $y$ direction with lateral gates,
the problem becomes unidimensional and the Hamiltonian 
for a one-dimensional wire reads\cite{StredaSeba}:
\begin{equation}
H_0 \, = \, \frac{p^2}{2 m^*} \sigma_0 \, - 
\, \frac{\alpha \langle {\cal E}_z \rangle}{\hbar} 
p_x \sigma_y \, + \, \frac{\epsilon_Z}{2 B } 
 {\vec B} . {\vec \sigma},
\end{equation}
where $m^*$ is the effective mass, $\langle {\cal E}_z \rangle$ is
the electric Rashba field perpendicular to the layer and $\alpha$ 
depends on the material used. 
${\vec B}$ is the magnetic field, $\epsilon_Z$ is the Zeeman energy.
$\sigma_0$ is identity matrix and 
${\vec \sigma} \, = \, (\sigma_x , \sigma_y , \sigma_z)$ the usual
Pauli matrices. The eigenvectors are thus products of plane waves 
times a spinor.
The eigenstates are:
\begin{equation}
{\rm E} \, = \, \frac{\hbar^2 }{2 m^*} \, \left\lbrack k_x^2 
 \pm 2 \sqrt{ \kappa_Z^4 + k_{\alpha}^2 k_x^2} \, \right\rbrack,
\end{equation}
where $\kappa_Z^2 \, =\, {m^* \over \hbar^2} (\epsilon_Z / 2)$ 
and $k_{\alpha} \, = \, {m^* \over \hbar^2} \alpha \langle {\cal E}_z \rangle$. 
The Rashba energy is defined
through $E_{\alpha} \, = \, {\hbar^2 \over 2 m^*} k_{\alpha}^2$.
The orientation of eigenvectors is such that
for $E \gg E_{\alpha}$, the spinor associated to the mode with
larger wave-vector is directed along
$\mid \uparrow \rangle_y$.
In the interval $\lbrack E_{\alpha} - \epsilon_Z / 2 \, , \, 
E_{\alpha} + \epsilon_Z / 2 \rbrack$, there is 
only one propagating mode (with two chiralities).
The other mode is evanescent. The dispersion
relation for the propagating mode is indicated in Fig. \ref{fig1}.
The existence of a pseudogap
in a given energy interval  has been used in Ref. \onlinecite{StredaSeba},
to propose a spin-filtering device, where a potential step of height
$V_1$ corresponding to a gate voltage permits to shoot in the middle 
of the pseudogap (Fig. \ref{fig1}).

Electron-electron interactions are taken into account, 
with $V_{int}(x)$, the Coulomb interaction potential.
Following Ref. \onlinecite{Matveev} which discusses
the effect of weak electron interactions in a single 
mode 1D wire, we use a Hartree-Fock approach followed
by a poor man Anderson renormalization in energy space.
The Dyson equation in Hartree-Fock approximation reads
\begin{eqnarray}         
{\vec \psi}_k(x) \, & = & \, {\vec \phi}_k(x) + \int dy \,
 G_k^r(x,y) \, V_H(y) {\vec \psi}_k(y) \, \nonumber \\
\,\! & \,\!  & + \int dy \int dz \, G_k^r(x,y) V_{ex}(y,z) {\vec \psi}_k(z),
\label{dyson}
\end{eqnarray}  
whith ${\vec \psi}_k$ (${\vec \phi}_k$)
the one-electron wave function 
in the presence (absence) of interactions, and 
$G_k^r(x,y)$ is the retarded Green function.
$V_H(x) \, = \, \int dy \, V_{int}(x-y) n(y)$,
is the Hartree potential,
with:
\begin{equation} 
n(y) =
\sum_{n=1}^2 \sum_{\mid q \mid <k_{F \, (n)}}
\mid {\vec \psi}_{q}^n(y) \mid^2~,
\end{equation}
the local density. 
Note that the sum is carried 
over the two modes ${\vec \psi}_{q}^{n}$
(propagating and evanescent), for all wave-vectors $q$
with energies smaller than $E_F$ (at $T=0$).
$k_{F \,(n)}$ are the wave-vectors of modes 
$n$ at the Fermi energy.
The exchange potential (Fock) is given by the matrix 
\begin{eqnarray} 
&& V_{ex}(x,y) =  - V_{int}(x-y) \nonumber \\
&& \times \, \sum_{n=1}^4 \sum_{\mid q \mid < k_{F\,(n)} }  \!\!\!
\begin{pmatrix}
\psi_{q , u}^{n \,\, ^*}(y) \psi_{q,u}^{n}(x) & 0 \cr 0 & 
\psi_{q,d}(y)^{n \,\, ^*} \psi_{q,d}^{n}(x)
\end{pmatrix}~,
\nonumber\\
\label{exchange} 
\end{eqnarray}
where $\psi_{q,u}^{n}$ and $\psi_{q,d}^{n}$ are the components
of ${\vec \psi}_q^n(x)$ in the
basis $\lbrace \mid \uparrow \rangle_z \, , \,
\mid \downarrow \rangle \rbrace_z$.
Assuming a finite range $d$ for the Coulomb interaction, 
we will consider in our perturbative calculation 
only the terms in Eq. (\ref{dyson}) which give the leading 
logarithmic divergence for the scattering coefficients 
at the Fermi level, as in Ref. \onlinecite{Matveev}.

In the case where the Fermi level lies in the middle of the 
pseudogap (Fig. \ref{fig1}), there 
are three scattering states. The first one corresponds to
injecting electrons from the left of the step
with an incident wave of wave-vector $k_+$ and group velocity
$v_{k_+}$, thus generating a reflected wave of amplitude $r_{1,1}$ and two 
transmitted waves, $t_{1,1} \sqrt{ {v_{k_+} \over v_{K_+}} } e^{i K_+ x}$
and $t_{1,2} \sqrt{ {v_{k_+} \over v_{K_-}} }  e^{i K_- x}$, 
$K_+$ being the larger wave-vector. The two other diffusion states
are obtained by injecting electrons from the right of the step with 
wave-vector $K_i$ ($i=+$ or $-$), thus generating reflected waves of
the form $r^{\prime}_{i,j} \sqrt{ {v_{K_i} \over v_{K_j} }} e^{i K_j x}$
and transmitted waves
$t^{\prime}_{i,1} \sqrt{ {v_{K_i} \over v_{k_+} }} e^{-i k_+ x}$, 
$j=+$ or $-$. The square root factors insure the unitarity
of the $S$ matrix \cite{LandauerButtikerMartin}. 

In the geometry of Fig. \ref{fig1}, the $S$ matrix thus takes the form
\begin{equation}
S \, = \,  \left( \begin{matrix}  r_{1,1}
  & t^{\prime}_{1,1} & t^{\prime}_{2,1} \cr 
t_{1,1}  &  r^{\prime}_{1,1} & r^{\prime}_{2,1} \cr
t_{1,2}  & r^{\prime}_{1,2}  &  r^{\prime}_{2,2} 
\end{matrix} \right).
\end{equation}
We introduce the physical parameter $\eta \equiv \sqrt{\epsilon_Z/E_{\alpha}}$
which is related to the ratio of the Zeeman energy to the Rashba energy.
The other relevant parameter is the ratio:
\begin{equation}
\Gamma = \sqrt{\frac{E_F}{E_{\alpha}}},
\end{equation}
$E_F$ is the energy of the Fermi level measured as in Fig. \ref{fig1}a. 
In our case, $E_F$ is connected to the height $V_1$ of the step. 
In order to inject electrons from the left, in the middle of the pseudogap, 
$E_F$ must be equal to $E_{\alpha} + V_1$.
Next, in order to have four propagating modes 
at the right of the step as in Fig. 1, $V_1$ must be larger
than $\epsilon_Z / 2$.  Thus, the minimum value of $\Gamma$
is $\sqrt{1+\epsilon_z/2E_\alpha}$. 
If $\eta \to 0$, $\Gamma \to 1$.
In the renormalization procedure,
following the usual procedure \cite{KaneFisher}, one then
rescales the bandwidth
from $D_0$, the real bandwidth, to $D$. Here $D_0$ is assumed to be 
larger than the step height $V_1$. 
One then obtains flow equations for the quantities
 $t_{1,1}$, $t_{1,2}$, $t^{\prime}_{1,1}$ and $t^{\prime}_{2,1}$.
It is more convenient to 
use ${\tilde t}_{i,j} = t_{i,j} \sqrt{{ v_{k_+} \over v_{K_j}}} $, 
and ${\tilde t}^{\prime}_{i,1} =
t^{\prime}_{i,1} \sqrt{ {v_{K_i} \over v_{k_+}}}$.
As an example, we quote the equation for ${\tilde t}_{1,1}$,
\begin{equation}
\frac{d {\tilde t}_{1,1} \,}{d \ln(\frac{D_0}{D}) } = 
-  (h {\overline v}_K)^{-1} 
\biggl\lbrack {\tilde t}_{1,1} \mid {\tilde r}^{\prime}_{2,1} \mid^2  
\Bigl( \frac{v_{K_+} }{ v_{K_-} } \Bigr)
  +  {\tilde t}_{1,2} r^{\prime}_{1,1} 
{\tilde r}^{\prime \, *}_{1,2}\biggr\rbrack
 {\cal J}^{\prime}_1, 
\end{equation}
where ${\overline v}_K  = {1 \over 2} \Bigl(v_{K_+} + v_{K_-} \Bigr)$.
The integral ${\cal J}^{\prime}_1$ is
 defined as ${\cal J}^{\prime}_1  =  {\cal I}^{\prime}_1 - {\cal I}^{\prime}_0$ where
\begin{eqnarray}
{\cal I}^{\prime}_0  & = & -{ v_{K_-} \over 2 {\overline v}_K } 
{\hat V}(K_{F+} + K_{F-} ) \nonumber \\
& &  \times  \ln \biggl\vert  
\Bigl\lbrace (K_+ +  K_- )  -  (K_{F+} +  \! K_{F-}) \Bigr\rbrace
 {d \over 2} \biggr\vert, \\
{\cal I}^{\prime}_1  & = &  -{ v_{K_-} \over 2 {\overline v}_K } 
{\hat V}\left( {(K_+ - K_{F+} ) + (K_- -  K_{F-}) \over 2} \right)  \nonumber \\
& &  \times \, \ln \biggl\vert  \Bigl\lbrace (K_+ - K_{F+} ) + (K_- - K_{F-} )
 \Bigr\rbrace { d \over 2}
 \biggr\vert~, 
\end{eqnarray}
where ${\hat V}(k)$ denotes the Fourier transform of $V_{int}$.
The renormalization equations for ${\tilde t}_{1,2}$, ${\tilde t}^{\prime}_{1,1}$ 
and ${\tilde t}^{\prime}_{2,1}$ have a similar structure.
${\cal J}^{\prime}_1>0$ for 
repulsive interactions.

Any  $S$ matrix
can be written as $exp(-i {\vec u} . {\vec \lambda})$
with ${\vec u}$ an 8-dimensional real vector  and 
the components of ${\vec \lambda}$ are real matrices ($SU(3)$). 
Since ${\tilde t}_{1,1}$, ${\tilde t}_{1,2}$, ${\tilde t}^{\prime}_{1,1}$ 
 and ${\tilde t}^{\prime}_{2,1}$ are generally complex numbers,
 our renormalization equations give 8 relations for real parameters 
and thus completely determine the flow of the $S$ matrix.
We now focus on the limit $\eta \to 0$ (small Zeeman coupling).
The $S$ matrix then simplifies and becomes a real and symmetric matrix:
\begin{equation}
S =  (\Gamma+1)^{-1}
\left( \begin{matrix}  {
(\Gamma-1)^2  \over \Gamma+1}
&  2 \sqrt{\Gamma}  &
- 2 \sqrt{\Gamma} {\Gamma-1 \over \Gamma+1 }\cr
 & & \cr 
2 \sqrt{\Gamma}  &
0 & \Gamma -1
 \cr & & \cr
- 2 \sqrt{\Gamma} {\Gamma-1 \over \Gamma+1 }
  &  \Gamma -1  & 
4 {\Gamma \over \Gamma+1} \cr 
\end{matrix} \right)~.
\label{smatrixgamma}
\end{equation}
It has a determinant of $-1$ instead of $+1$ because the number 
of propagating modes is not the same on each side.
The $S$ matrix stays
real symmetric under the renormalization group flow and thus,
only two real equations for $t_{1,1}$ and $t_{1,2}$ are needed.
 
When injecting electrons from the left of the step 
(exactly in the middle of the pseudogap as in 
Ref. \onlinecite{StredaSeba}),
one defines the polarization of the outgoing 
beam as:
\begin{equation}
p \, = \, {\mid t_{1,1} \mid^2 - \mid t_{1,2} \mid^2 \over 
 \mid t_{1,1} \mid^2 + \mid t_{1,2} \mid^2 }.
\end{equation}

The value $p=0$ corresponds to a totally unpolarized beam
(experimentally not desirable). 
The value $p = 1$ corresponds to a perfectly polarized outgoing
beam. According to Eq. \ref{smatrixgamma}, $p$ is simply $\Gamma^{-1}$. 
It tends to zero as the height of the step gets large 
with respect to the Rashba energy. The propagating modes
are either oriented along the $\mid \uparrow \rangle_y$ 
or along the reverse direction.

\begin{figure}[h] 
\epsfxsize 7 cm  
\epsfysize 4.5 cm
\centerline{\epsffile{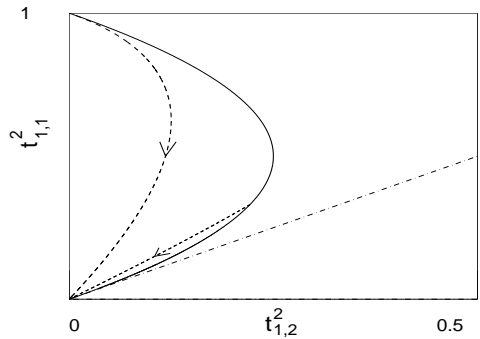}}
\caption{Solid line: location of the initial points. Upper dashed line:
renormalization group flow for $\Gamma=1$. The starting point
is at $t_{1,1}=1$. Lower dashed line: renormalization group trajectory
for $\Gamma = 10.$ The dashed-dotted line is simply the first
diagonal, corresponding to totally unpolarized beams 
and reaches $t_{1,1}^2=1$ at $t_{1,2}^2=1$.\label{fig2}}
\end{figure}

Let $l \, = \, \ln \Bigl( {D_0 \over D} \Bigr)$ be the renormalization parameter 
and ${\rm A} \, = \,  {\cal J}^{\prime}_1 / (2 \pi \hbar {\bar v}_K) $.
It is easier to express the renormalization group equations in terms of the
 variables $x \, = \, t_{1,1}^2 + t_{1,2}^2$
 and $y \, = \, t_{1,1}^2 - t_{1,2}^2$.
They read:
\begin{eqnarray}
{dx \over dl} & = & - {\rm A} (x^2 - y^2) \sqrt{1-x}   
\Bigl( 1-\sqrt{1-x} \Bigr)~, \label{rgx}\\
{dy \over dl} & = &  - {1 \over 2} {\rm A} (x^2 - y^2) 
y~.\label{rgy}  
\end{eqnarray}
There are only 3 fixed points. The stable fixed point $x=0,y=0$ 
corresponds to a perfectly reflecting step; both
$t_{1,1}$ and $t_{1,2}$ are zero. The unstable fixed point $x=y=1$
corresponds to $t_{1,1} =1$ and $t_{1,2}=0$, i.e. perfect
transmission with no mode conversion. The second unstable fixed 
point $x=1,y=-1$ corresponds to $t_{1,2}=-1$ and $t_{1,1}=0$ 
(complete mode conversion).
When we have no electron-electron interactions, the location of the 
points where one starts the renormalization procedure, in the plane 
$(t_{1,2}^2,t_{1,1}^2)$, depends
on one parameter only, $\Gamma$. Thus, the ensemble of points
where one starts is a curve in this plane, which is represented in Fig. \ref{fig2} 
 as a solid line. In addition, the behavior of some flow trajectories
is depicted in this figure. 

We describe what happens first, 
if the height of the step is small with respect to the Rashba 
energy, second, in the reverse situation.
\begin{figure}[h] 
\epsfxsize 7. cm
\epsffile{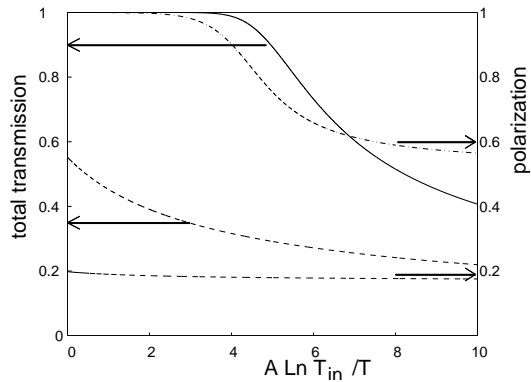}
\caption{Total transmission and polarization as a function
of the logarithm of the temperature. The two curves marked 
with left arrows represent the total transmission for $\Gamma=1.01$ (solid line) and 
$\Gamma = 10$ (dashed line).
The two other curves represent the spin polarization 
for $\Gamma = 1.01$ (dash-dotted line)
and $\Gamma = 10.$ (dashed line) respectively.
\label{fig3}}
\end{figure} 
In the first case, $\Gamma$ is barely larger than $1$ ($1$ ideally).
We then start from an initial temperature (bandwidth) $T_{in} = D_0/k_B$, where
$t_{1,1}$ is close to $1$ and $t_{1,2}$ close to zero, 
which means a perfect polarization of the outgoing beam.
The point moves first in a direction such that the polarization 
diminishes significantly but the total transmission remains 
approximately constant. Then, the curve bends downwards and
eventually moves towards the origin (no transmission). 
The polarization goes to a constant value.

In the second case (not desired in practice), $\Gamma$ is large, 
the initial polarization is already low (equal to 
$2/\Gamma$) and still decreases towards a constant value.
The transmission goes quickly to zero.  

The renormalization procedure must be interrupted at some stage
given by $D \simeq k_B T$. In the case where $\Gamma$ is close to $1$
(small potential step), Eqs. (\ref{rgx}) and (\ref{rgy}) show that the total 
transmission first decreases very slowly as
$1- r_{in}^2 - 8 r_{in}^2 {\mathcal A} \ln^2 \bigl( T / T_{in}  \bigr)$,
where $r_{in}$ is the initial reflexion coefficient. $r_{in}$ goes 
to zero as $\Gamma$ goes to $1$. 
The polarization decreases as $ \ln(T_{in}/T)$.
For lower temperatures, the polarization goes to a constant 
$\Gamma$ dependent value, which is 
equal to $1/2$ for $\Gamma=1$. The total transmission 
goes to zero, asymptotically like
$\ln 
(T_{in} / T )^{-1\over 2}
$, but this regime
is only attained for unrealistic values of temperatures.

A characterization of such a spin filter requires the 
comparison of both 
the total transmission coefficient,
together with the evolution of the polarization under 
renormalization group flow. This information is illustrated 
in Fig. \ref{fig3} where both quantities are plotted as a 
function of the logarithm of the inverse temperature. 
For $\Gamma$ close to $1$, the total transmission stays constant 
and close to unity until the temperature is dropped 
by several orders of magnitude, and 
then decreases in a monotonous fashion. 
At the same time, the polarization drops faster than the total transmission, 
signifying that the quality of the spin filtering effect is polluted 
by electronic interactions before the total transmission 
is truly affected. 
For large $\Gamma$, the total transmission 
first decreases monotonously 
(linear behavior in $\ln T$) starting 
from the initial bandwidth, then giving place to 
a slower decrease ($(\ln T)^{- {1 \over 2}}$). 
The polarization stays approximately constant 
in this case, but at a 
deceptively low value of $0.2$.  
%
%


A useful way to quantify electron interactions
is to compute the Luttinger parameter $g$ which is expected for 
a gated heterostructure.
To be specific we consider the geometry of Ref. \onlinecite{yacobi}, 
where the Coulomb interaction in the 1D channel is screened by
the few transverse modes in the wire and by the proximity of the 
2DEG and gates: $V(r)\simeq (e^2/4\pi \epsilon_0\epsilon)e^{-r/\lambda_s}/r$
($\lambda_s\sim 100 nm$, a fraction of the width of the wire, 
$W\sim 20 nm$).
Averaging over the lateral dimensions of the wire, one obtains 
an effective one-dimensional potential.
The Luttinger parameter
$g$ is then related to the zero-momentum Fourier 
transform of this potential
$g=(1+4\tilde{V}(0)/\hbar v_F)^{-1/2}$.
$g$ increases with the ratio $W/\lambda_s$.  
Taking $v_F\sim 10^6 m/s$ one obtains $g\simeq 0.69$, 
which is remarkably close to the value of Ref. \onlinecite{yacobi}
It is also reasonably close to the non-interacting
value $1$. 
The materials used in Ref. \onlinecite{yacobi} differ
from the existing Rashba devices \cite{nitta}, but 
the typical parameters are comparable.

To summarize, we have looked at the effect of weak 
electron electron interactions in one dimensional ballistic 
quantum wires under the combined Rashba and Zeeman effects. 
At the single electron level this device has the 
advantage of working as a spin filter. We have characterized 
the influence of electron electron interactions in this same 
device. We found that in the most relevant case 
($E_F-E_\alpha \ll 1$) where the total transmission remains
close to unity for a range of bandwidths,  
the quality of spin filtering properties decreases  
substantially.
Although the present work deals with a sharp step, 
the present approach can be extended to steps whose extension 
is much larger than the Fermi wave length, using WKB-type approximations.
Transmission through the step is likely to be enhanced in this case, 
but the tendency for interactions to spoil filtering
will remain valid.   
%
%
Also, this single electron picture can be further 
complicated in practice:
for the case where two spin orientations are present
(on the right hand side of the step),  
it was shown \cite{FribourgKarlsruhe} that the combined effect
of spin orbit  coupling and strong electron electron 
interactions provide some limitations to the Luttinger liquid 
(metallic) picture, giving rise to spin or charge 
density wave behavior instead.
Nevertheless,    
the present perturbative treatment of the Coulomb interaction 
is justified here because the one dimensional wire is 
surrounded by nearby metallic gates in order to implement 
the Rashba effect.

We thank P. St{\v r}eda and F. Hekking for
useful discussions and comments. This work 
is supported by a CNRS A.C. 
Nanosciences grant.

\end{document}